\documentclass[a4paper,10pt,review]{elsarticle}

\usepackage{multicol}
\usepackage{color}
\usepackage{xspace}
\usepackage{pdfwidgets}
\usepackage{enumerate}

\def\ttdefault{cmtt}

\usepackage{amsmath}
\usepackage{mathptmx}
\usepackage{graphicx}

\usepackage[english]{babel}

\usepackage{url,hyperref,lineno,microtype,siunitx}
\usepackage{xcolor}
\usepackage[normalem]{ulem}
\definecolor{MyDarkGreen}{rgb}{0,0.8,0.0}
\definecolor{MyDarkBlue}{rgb}{0,0,0.8}
\definecolor{MyDarkRed}{rgb}{0.6,0,0.0}

\begin{document}

\title{Simulation, visualization and analysis tools for pattern recognition assessment with spiking neuronal networks}

\author[1]{Sergio E. Galindo\fnref{fn1}}
\ead{sergio.galindo@urjc.es}

\author[2]{Pablo Toharia\fnref{fn2,fn3}}
\ead{ptoharia@fi.upm.es}

\author[1]{Oscar D. Robles\fnref{fn1,fn3}}
\ead{oscardavid.robles@urjc.es}

\author[3]{Eduardo Ros\fnref{fn4}}
\ead{eros@ugr.es}

\author[1]{Luis Pastor\fnref{fn1,fn3}}
\ead{luis.pastor@urjc.es}

\author[3]{Jes\'us A. Garrido\corref{cor1}\fnref{fn4}}
\ead{jesusgarrido@ugr.es}

\cortext[cor1]{Corresponding author}
\fntext[fn1]{Universidad Rey Juan Carlos, Madrid, Spain}
\fntext[fn2]{Universidad Polit\'ecnica de Madrid, Spain}
\fntext[fn3]{Center for Computational Simulation, Universidad Polit\'ecnica de Madrid, Spain}
\fntext[fn4]{Universidad de Granada, Spain}

\address[1]{C/ Tulip\'an, S/N, 28933, M\'ostoles, Madrid, Spain}
\address[2]{C/ de los Ciruelos, 28660 Boadilla del Monte, Madrid, Spain}
\address[3]{C/ Periodista Daniel Saucedo Aranda, S/N, E-18071, Granada, Spain}

\begin{abstract}
Computational modeling is becoming a widely used methodology in modern neuroscience. However, as the complexity of the phenomena under study increases, the analysis of the results emerging from the simulations concomitantly becomes more challenging. In particular, the configuration and validation of brain circuits involving learning often require the processing of large amounts of action potentials and their comparison to the stimulation being presented to the input of the system. In this study we present a systematic work-flow for the configuration of spiking-neuronal-network-based learning systems including evolutionary algorithms for information transmission optimization, advanced visualization tools for the validation of the best suitable configuration and customized scripts for final quantitative evaluation of the learning capabilities. By integrating both grouped action potential information and stimulation-related events, the proposed visualization framework provides qualitatively assessment of the evolution of the learning process in the simulation under study. The proposed work-flow has been used to study how receptive fields emerge in a network of inhibitory interneurons with excitatory and inhibitory spike-timing dependent plasticity when it is exposed to repetitive and partially overlapped stimulation patterns. According to our results, the output population reliably detected the presence of the stimulation patterns, even when the fan-in ratio of the interneurons was considerably restricted.
\end{abstract}


\maketitle


\section{Introduction and motivation}
Computational modeling of brain circuits represents an affordable, versatile and powerful neuroscience tool for studying brain function, and it has become a key component of some brain-studying initiatives such as the Human Brain Project \cite{markram2012}. By using computational models, it can be determined the influence of several biological phenomena with different level of detail and how their interaction results in advanced behavior. However, as the phenomenon being modeled and the network under study become more complex, the tools required for fully understanding and validating the underlying biological principles become more challenging as well.

Besides the usage of computational neuroscience models to study information processing in the brain, spiking neuronal networks are also called on to revolve deep neuronal networks by providing massively parallel, energetically efficient learning systems which naturally represent spatiotemporal knowledge \cite{Kasabov2019}. Deep and convolutional neuronal networks have recently succeeded in extracting knowledge from complex processes based on multidimensional and multimodal data, representing the current state-of-the-art in artificial intelligence methods. Classical deep neuronal networks represent data as a series of vectors. These data are the substrate for extracting information during the learning procedure \cite{Bishop2006}. However, the main limitation of these networks is that they do not represent time directly in the model \cite{Lecun2015}. On the contrary, spiking neuronal networks encode spatiotemporal information as trains of spikes over time (ranging from milliseconds to tens of milliseconds) in a similar way as the brain does it. However, the development of these kinds of learning systems requires a range of tools that facilitate the understanding of the information representation and the validation of the learning capabilities.

In particular, information encoding in neuronal populations and the evolution of the sensorial representation in learning systems are open topics where computational models play a pivotal role. Traditionally, the analysis of the simulations within computational neuroscience studies relies on home-made scripts to extract specific quantitative estimators and complex figures for qualitative evaluation of the simulation output. Recently, software packages have emerged aiming to facilitate the analysis and processing of neuronal activity \cite{Sobolev2014} in both simulated and experimental frameworks. Similarly, other software packages, including StackViz and SimPart, have been proposed to allow researchers fast visual analysis of the network activity\cite{galindo:2016}.

The work by Kasinski et al. \cite{Kasinski:2009} provides a 3D visualization for the analysis of dynamical processes in spiking neuronal networks, including neurons, inter-neuronal connections and spikes. Then, a set of graph windows display the firing times or membrane potential of certain selected neurons. Nowke and colleagues presented VisNEST\cite{nowke:2013,nowke:2015}, a framework for the visualization of activity of multi-area network models, including spiking events and the mean firing rate. In a different view a dynamic 3D graph shows the spike exchange between different cortical areas, in combination with classic charts used to represent spatial information. The Visualization of Layer Activity (VIOLA)\cite{senk:2017} should also be mentioned, a web-based application that shows spiking activity data from massively parallel neurophysiological data, exploring spatiotemporal features of the neuronal activity as well as a quantitative analysis of specific aspects of the data. However, all these approaches lack support for the analysis of information transmission and the emergence of populations responsive to the presentations of specific patterns. This represents a challenging difficulty in learning-system development that needs to be overcome by means of the users' code.

The development of novel population activity visualization tools may open up groundbreaking possibilities for the study of learning evolution in spiking neuronal circuits. Brain networks are composed of populations of excitatory and inhibitory neurons that transmit complex action potential (spike) patterns through their synapses. Synapses can persistently modify their efficiency in an activity dependent manner, providing the biological substrate for learning and memory. In particular, spike-timing dependent plasticity (STDP) represents a family of synaptic mechanisms in which the long-term potentiation or depression depends on the temporal shift between presynaptic and postsynaptic spike occurrences \cite{bi2001}. STDP mechanisms have been recorded in both excitatory \cite{bi2001,damour2015} and inhibitory synapses \cite{haas2006,damour2015}, and theoretical studies have demonstrated the potential of this rule for learning and detecting spike patterns in several application frameworks including, but not restricted to, spatiotemporal spike patterns \cite{masquelier2009spike}, current-based analog patterns \cite{masquelier2009current,Garrido2016} and visual features \cite{masquelier2007}.

One of the main limitations of STDP-based learning systems is the high number of afferent synapses that is required in order to achieve accurate pattern detection. A previous simulation study showed that repetitive spike patterns need to account for 10 percent of the full input population to become reliably detected \cite{masquelier2009current}. However, whether the total population size and the presence of homeostatic mechanisms in the neurons (such as the firing threshold adaptability) affect the capability of the network to learn and detect input patterns still remains elusive.

\begin{figure*}[htb]
    \includegraphics[width=1.00\textwidth]{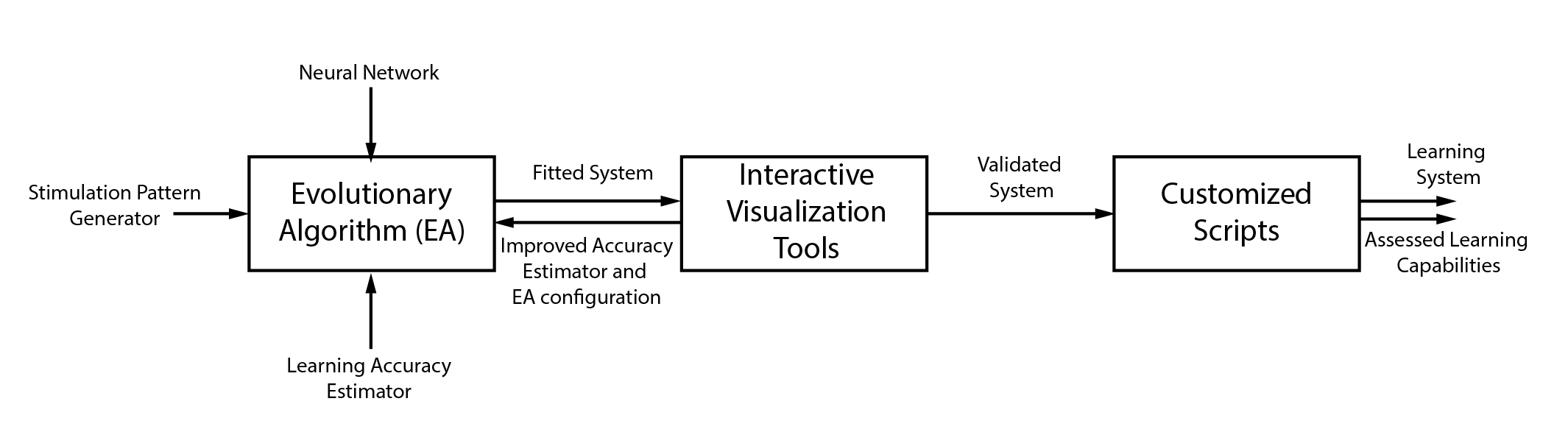}
    \caption{Schematic representation of the presented work-flow for development of learning systems with spiking neuronal networks. The evolutionary algorithm (EA) adjusts the neuronal network parameters by running the simulation with the pattern generated by the stimulation pattern generator. The goodness of the configuration is estimated using the learning accuracy estimator (fitness function). This measure allows the selection of the most promising individual for further exploration. The resulting individuals emerging from the EA will be visually validated with the visualization tools (StackViz and SimPart), ensuring that the individual behaves as expected during the simulation. If this is not the case, the visualization of the activity allows us to refine the accuracy estimator and the EA exploration parameters. Finally, the performance of the validated learning system will be assessed with customized scripts.}
  \label{fig:work-flow}
\end{figure*}

In this article, StackViz and SimPart visualization tools have been further developed to assess the evolution of information transmission during the learning process. These tools, together with evolutionary algorithms (EA) for network parameter tuning and custom scripts for quantitative evaluation of learning represent an integrative work-flow for developing unsupervised learning systems with spiking neurons. The work-flow proposed in this theoretical study includes the following stages (Fig. \ref{fig:work-flow}):
\begin{enumerate}
	\item Evolutionary algorithms for network parameter optimization. Each individual evaluated the network with a single parameter configuration and the solutions were ranked according to different information transmission estimators.
	\item The selected individual simulations were visualized using SimPart. It allowed the validation of the network connectivity and the stimulation protocols by observing the spatial distribution of action potentials.
	\item Simultaneously, the individual simulations were visualized using StackViz. It enabled fast quantitative assessment of the different sub-populations emerging from the learning process and how their activity was correlated with the presence of the stimulation patterns.
	\item Finally, customized Python scripts analyzed the simulation data to obtain specific quantification of the learning accuracy, such as the $UC$ and $LA$ indexes.
\end{enumerate}

The visualization and analysis work-flow has been used to study how receptive fields emerge in a learning paradigm involving theta-band frequency oscillations, STDP and inhibitory interneurons. The proposed learning system has shown accurate learning and the detection of multiple repetitive overlapping stimulation patterns, as validated with the visualization framework. Finally, the influence of the interneuron fan-in ratio in the learning accuracy has been characterized.

The rest of the paper is organized as follows: Section \ref{sec:2-description} describes the visualization framework and the improvements made for this study; Section \ref{sec:3-application-case} further elaborates the application case from the computational modeling perspective, defining the network under evaluation and learning protocol being used; Section \ref{sec:4-results} analyses the results obtained from the simulations and validate them with the visualization tools and finally, Section \ref{sec:5-discussion} discusses the results as well as the conclusions of this research.

\section{Visualization and analysis framework: ViSimpl} \label{sec:2-description}
\subsection{Visualization framework description}

\begin{figure*}
	\includegraphics[width=1.00\textwidth]{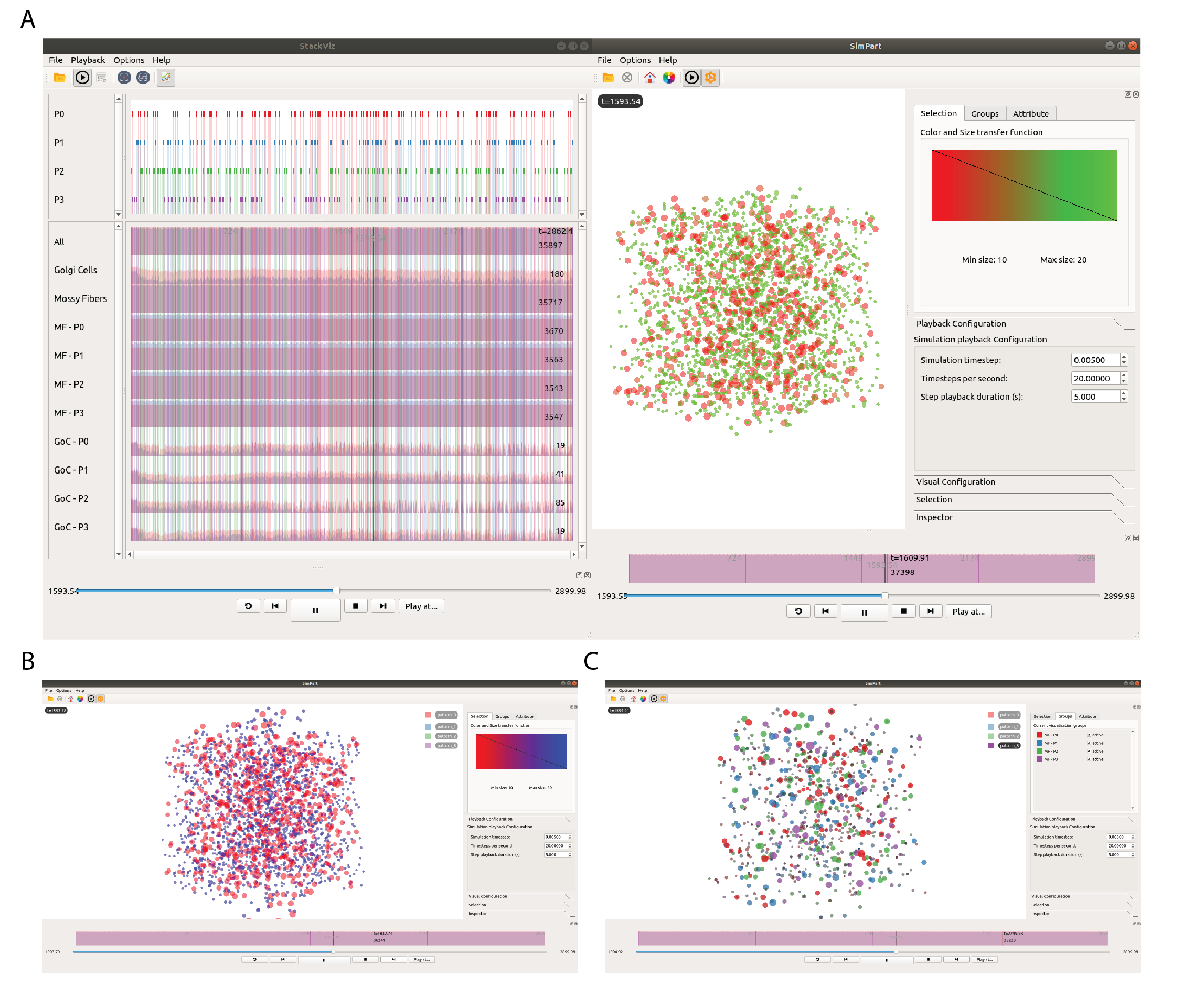}
	\caption{Graphical User Interface of the visualization framework tools. (A) On the left, StackViz displays a set of activity plots overlapped with vertical bars representing the time events (presentation of patterns) of the actual simulation. On the right, SimPart shows the current activation state of the involved neurons classified by groups and indicating whether input events are activated at the current time. Both applications are synchronized, in this case sharing the current playback status, the present subsets and the colors being used to represent pattern activation on both sides. (B) Activity-based color variation in SimPart. The configured transfer function (color, size and transparency) is applied depending on the actual state of each neuron after spiking, transiting from red to blue whenever activated. (C) Functional-type-based color variation in SimPart. Each neuron group has a different transfer function that varies in brightness according to the current activation state of each neuron.}
	\label{fig:visimpl-all-views}
\end{figure*}

The visualization and analysis framework used in this work was ViSimpl \cite{galindo:2016}. As described in a previous paper, it is composed of different applications designed to perform both quantitative and qualitative analysis of brain simulation data. From ViSimpl framework, StackViz and Simpart were specifically used to perform temporal analysis and to analyze the data from a spatio-temporal perspective respectively. Both tools were synchronized to operate together using ZeroEQ \cite{zeroeq} for exchanging playback operations and sets of elements \cite{galindo:2016}.

StackViz is an application for the interactive analysis and comparison of time-varying series representations, in this case, simulation activity histograms. This tool allows interactive exploration of the data by comparing, performing ``focus + context" operations, zooming and even controlling the visualization playback being run on SimPart. In a previous study the plot stacking feature was introduced as a way of visually analyzing and comparing the distribution of activity between different subsets of neurons throughout the simulation. When a selection of elements is generated or received by StackViz a new plot is stacked at the bottom of the display area showing the activity histogram performed by these elements. The plots show two different representations of the same activity for two comparison purposes: local normalization (displayed in blue) and global normalization (displayed in red). The former normalizes the graph according to the local maximum of the activity distribution which establishes the context to compare activation peaks for the same subset, whilst the latter normalizes the graph using the maximum of all the activity of the neurons, allowing users to perform comparisons amongst different subsets. Different visualization scales (namely, linear or logarithmic) may be selected for each normalization type. Unless otherwise stated, a linear scale for local normalization and a logarithmic scale for global normalization were used. 

SimPart is a tool designed for analyzing time-varying series over a three-dimensional environment, representing each neuron with particles as a translucent point or sphere at its given 3D position. In order to model neuronal representation throughout the simulation, a transfer function (color, size and transparency) is used for transiting from an activation state when spiking until they reach a rest state, changing their visual aspect during a specific time window. Visualization may be interactively explored by using the simulation playback, a feature introduced in the author's previous study, which allows the user to effectively control through operations such as pause/resume, jump to an arbitrary part of the simulation, restart or loop. The user has the capability of dynamically tuning the simulation playback (timestep duration and steps per second) as well as the visualization parameters (transfer function, transparency mode, activation duration) at any time.

In this study the simulation playback was improved with the ``step by step" capability, a feature for representing the given timespan within a certain time window with a higher time resolution than the normal playback mode for a more detailed activity exploration. This is defined by the increment of simulation time (timestep duration) and the desired length of playback in seconds. Interestingly, this new feature allows the user to clearly visualize the neurons that become active in response to the presentation of one or several stimulation patterns (see Section \ref{subsec:results-multiple-pattern}).

The underlying process to visualize spike trains was also improved for this study; the problem relies on the sparse nature of action potentials, as jumping to an arbitrary time stamp requires recreating the previous status at the specific start time for all the active neurons in order to apply the new frame time deltas, considering both the previous state (based on configured decay) and the current frame elapsed time. Therefore, visualization is correct at any arbitrary time when the user jumps between parts of the simulation.

In the following sections several new features will be introduced. These features were designed to provide capabilities for a more exhaustive analysis of neuronal simulation data and specifically for the visual correlation of input patterns and neuronal activity, a task which plays a key role in the use case objective of this study.

\subsection{Input Events}

The use case described in this study involves the capability of spiking neuronal networks to evolve receptive fields based on correlated inputs (repetitive patterns). The spatial input patterns (between 1 and 4 depending on the specific experiment) were randomly chosen at the beginning of the simulation and repeated frequently (Fig. \ref{fig:visimpl-all-views}A) beyond the background noise signal. Therefore, the presence of input patterns might be scattered all over the simulation time. This segmentation of input signals was modeled as discontinuous sets of input activation times for a determined set of neurons that occur throughout the simulation.

As this article aimed to observe the influence of the input patterns in the activity of the neurons, a way to visually analyze whether the characterized set of neurons is responding to the expected input or not was modeled. Thus, users may benefit from examining interesting parts of the simulation with hints of when and where activity may be present. As a result, input events were represented as part of the plot-stacking visualization (StackViz) and the spatio-temporal one (SimPart).

In StackViz events were expressed as a series of time lapses represented as semi-transparent rectangles covering the area where the event was activated, regardless of the input value. Once they were loaded on the application, a new widget appeared at the top part of the application showing stacked time events and their evolution was rendered using the same time scale as the activity plots. This feature was represented by overlapping the horizontal plots displaying the activity of a particular neuron set with vertical bars that represented the periods of time where the inputs were active. These vertical bars were rendered using different semi-transparent colors for each, so it was easy to distinguish between time events as shown in Figure \ref{fig:visimpl-all-views}A. Transparency allowed the representation of these events to be overlapped with one another and also with the activity plots, so the user may visually match the input with the activity performed by the network throughout the simulation.

SimPart represented events as a series of visual indicators (one per event set) that were highlighted whenever the event was activated. These visual indicators contain a label with the name of the event set and were displayed using different colors that matched the ones used in StackViz. The user may then relate a visually perceived spatial pattern of activity to the activation of an input.

Based on these events and the process that it is further described in Section \ref{subsec:selectivity-analysis}, the neuron sets were specified considering their functional classification (such as those neurons whose activity is the most representative of the stimulation pattern presentation) for the later disaggregated exploration of their activity. This visual analysis leads to a faster validation of the network learning capabilities, providing visual comparison between the spiking activity of arbitrary subpopulations and the presence of specific stimulation patterns in the input of the network under study. In addition, these subsets somehow related to the event activity were directly translated into visual groups as explained in the following section.

\subsection{Visualization groups}

Once the calculation of event-correlated subsets was performed, the capability of analyzing the activity of these groups independently from the spatio-temporal perspective arose as something desirable and useful. The color mapping described in the author's previous paper \cite{galindo:2016} established color variations depending on the current state of the neuron activation, but it proved to be inefficient to simultaneously analyze the activity of diverse subsets. This problem was originated by the absence of a mechanism that separates groups, avoided color mixing and reduced visual cluttering as can be seen in Figure \ref{fig:visimpl-all-views}B.

To overcome these problems the definition of groups was introduced in SimPart, that led to the capacity of disaggregating the data, represented using different color transfer functions in order to differentiate the activity of diverse sets of neurons, taking into account their functional type. The user may select a set of neurons using their GIDs and create a visual group, assigning to it certain color and size transfer functions. From that moment onwards, this selection is shown as a group distinguished from others by its color mapping, which consists of a monotonic brightness variation from a base color to a darker version of it. These visual groups may be interactively added, removed, activated and deactivated whenever the user wants to hide some detail from the simulation visualization.

Grouping was used to visually identify different subsets present in StackViz that were interesting to be rendered together but separating their representation. This new feature allowed visual assessment of how different neuron subsets respond to pattern activation. A visual group may be created for a given subset of input fibers, as well as the interneurons whose activity is highly correlated with the presence of input patterns or any other arbitrary selection. As the default mode, these groups changed color based on their corresponding levels of activity. Figure \ref{fig:visimpl-all-views}C illustrates how reducing the visual clutter by distributing neurons between functional groups facilitates observing how receptive fields emerge as the simulation time advances.

\section{Application case. Receptive fields building process through eSTDP and iSTDP} \label{sec:3-application-case}
Aiming to validate the applicability of the presented visualization tools into a theoretical neuroscience study previous research has been further refined. Our previous study explored the way excitatory spike-timing dependent plasticity (eSTDP), inhibitory spike-timing dependent plasticity (iSTDP) and input oscillatory activity allowed current patterns to be automatically recognized in recurrent neuronal networks of interneurons \cite{Garrido2016}. The combination of all these mechanisms resulted in the emergence of receptive fields in the excitatory (afferent) connections of the network. In this study, this framework have been further extended to explore how the input connectivity ratio (e.g. the number of afferent fibers) of the interneurons affects the learning capabilities of the network.

This section will describe the computational model, the optimization algorithm for parameter tuning and the estimators being used to assess the learning capabilities of the system.

\subsection{Description of the recurrent neuronal network}
\label{subsec:application-neuron-model}

\begin{figure*}[htb]
    \includegraphics[width=1.00\textwidth]{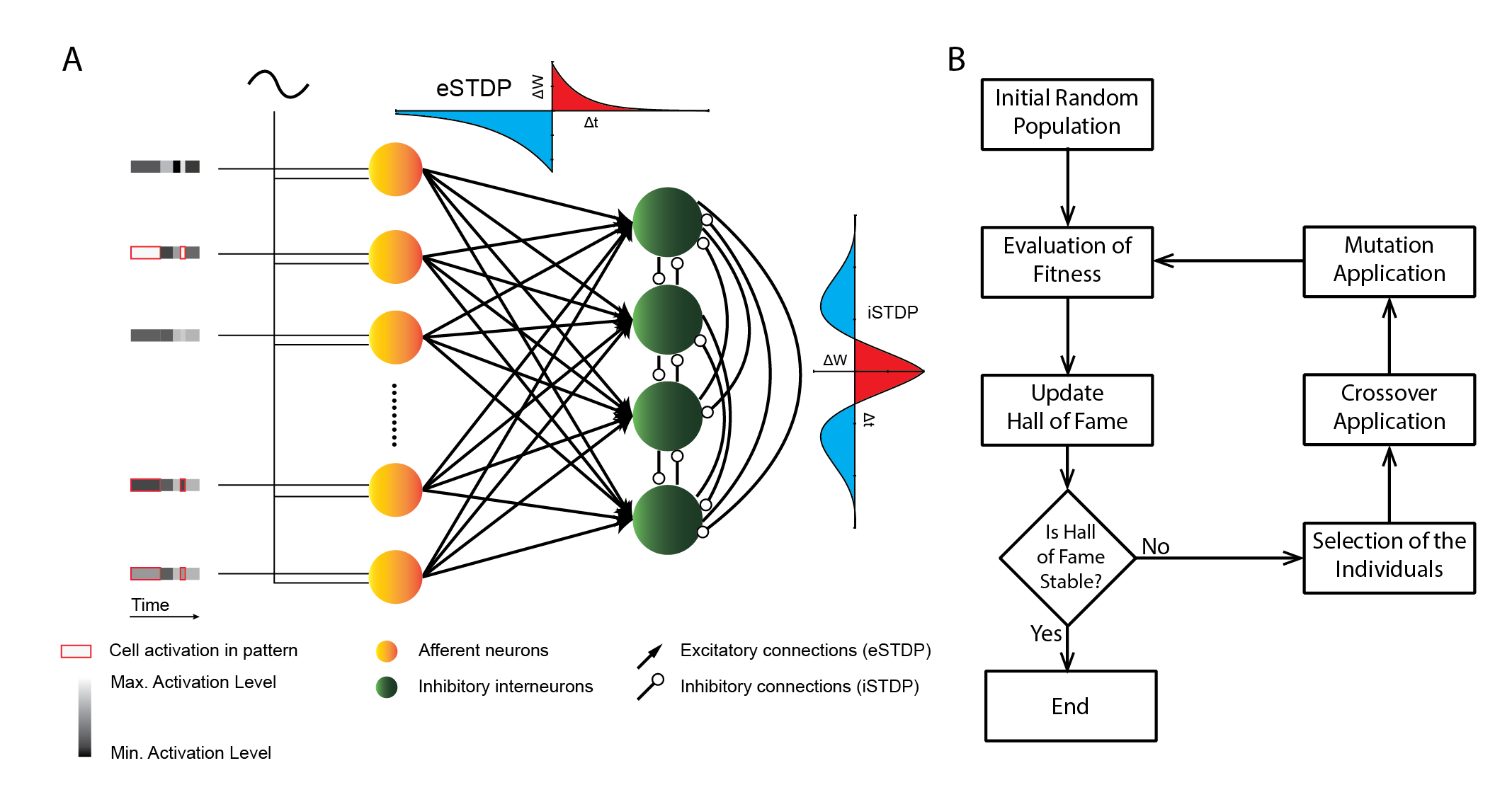}
    \caption{(A) Schematic of the simulated network including the stimulation, pattern generation protocol and plasticity mechanisms. The neuronal network is composed of 2000 excitatory neurons (yellow circles) which receive external stimulation composed of a carrier 8-Hz oscillatory signal in addition to randomly generated input combinations (embedding the presented patterns). These neurons make excitatory synapse (with eSTDP, -arrow-ending connectors-) with the inhibitory interneurons (green circles). The interneurons laterally inhibit each other (with iSTDP -circle-ending connectors-). The excitatory and inhibitory STDP learning rules have been plotted, respectively, at the top and right of the synapse. These functions regulate weight change as a function of the time difference between the presynaptic and postsynaptic spike. The blue areas represent long-term depression (LTD) whilst the red areas indicate long-term potentiation (LTP). (B) Flowchart of the evolutionary algorithm implemented for the optimization of the network configuration.}
  \label{fig:network-model}
\end{figure*}

The network model presented here was inspired by the topological structure of the cerebellar input layer. A network of spiking neurons was created containing mossy fibers (afferent neurons) and Golgi cells (inhibitory interneurons) (Fig. \ref{fig:network-model}A). The neuronal density was supported by histological data in the cerebellar granular layer (Table \ref{tab:number_of_neurons}). Specifically, the activity of the interneurons (namely, the Golgi cells) existing in a cube \SI{2}{\micro\metre}s was simulated on each side of the cerebellar granular layer. The neurons were placed in the cube according to a uniform distribution only for visualization purposes, as the position had no influence on the connectivity of the network. The network model was implemented using NEST 2.14 simulator \cite{nest_2_14}.

\begin{table}
\centering
	\caption{Network element density and number of neurons used in the network simulation}       
	\begin{tabular}{llll}
		\hline\noalign{\smallskip}
		Neuron layer & Density & Neurons & Reference\\ 
		\noalign{\smallskip}\hline\noalign{\smallskip}
		Afferent neurons & \num{3.0d5}\si{mm^{-3}} & 2400 & \cite{Jakab1988}\\
		Interneurons & \num{9.0d3}\si{mm^{-3}} & 72 & \cite{Korbo1993}\\
		\noalign{\smallskip}\hline
	\end{tabular}
  \label{tab:number_of_neurons}
\end{table}

Once the neurons were located in their positions the convergence ratio was explored to assess the impact on the learning capabilities of the network. Each interneuron was connected to a fixed number of excitatory sensorial afferents ranging between 50 and 2000. The connectivity was made following uniform random distributions to avoid any undesired effect in the limits of the cube. In addition, each interneuron received lateral inhibition from each other.

A more detailed description of the neuron and synaptic models and all the equations governing the dynamics have been included as supplementary information. The afferent neurons were modeled as current-based versions of the Leaky Integrate-and-Fire (LIF) neuron model while the interneurons used conductance-based versions of the same neuron type. In order to ensure appropriate average firing range independently of the particular weight configuration, the impact of equipping the interneurons with an adaptation mechanism for the firing threshold \cite{Huang2016} has been explored, according to the following equation:

\begin{equation}
\label{eq:ad_thr_diff}
\frac{dV_{th}}{dt} = -\frac{V_{th}-E_{inh}}{\tau_{th}}
\end{equation}

where $V_{th}$ represents the firing threshold, $E_{inh}$ is the reversal potential of the inhibitory synapse and $\tau_{th}$ is the adaptive threshold time constant. According to this equation, in absence of action potentials, the threshold progressively decreases towards the minimum value (the same as the inhibitory reversal potential), facilitating neuron firing. Similarly, when a spike is elicited, the firing threshold is increased a fixed amount proportional to the constant $C_{th}$ as indicated in Eq. \ref{eq:ad_thr}, making neuron firing more sparse.

\begin{equation}
\label{eq:ad_thr}
\delta V_{th} = \frac{C_{th}}{\tau_{th}}
\end{equation}

The neuron model parameters were set as indicated in Table \ref{tab:neuron_parameters}. The sensorial afferent parameters (and stimulation current values) were conveniently chosen to generate between 1 and 3 spikes per oscillatory cycle. These activity rates ensured homogeneous evolution of the synaptic weights connecting all the afferents. Interneuron electrophysiological parameters were set to values previously reported in the literature for GoC neurons \cite{Garrido2013a}. Finally, threshold adaptation parameters were set after preliminary tests aiming to stabilize average firing rates in the range of 2Hz ($C_{th}$) and firing rate evolution due to adaptive threshold several orders of magnitude slower than the synaptic plasticity ($\tau_{th}$).

\begin{table}
\centering
	\caption{Excitatory afferent and inhibitory interneuron model parameters. Note that those parameters that were not used are indicated with a hyphen(-).}       
	\begin{tabular}{lllll}
		\hline\noalign{\smallskip}
		Parameter & Afferents & Interneurons\\ 
		\noalign{\smallskip}\hline\noalign{\smallskip}
		$C_m$ &  \SI{2.0}{\nano\farad} & \SI{50}{\pico\farad} \\
		$G_{leak}$ &  \SI{500}{\nano\siemens} & \SI{3}{\nano\siemens} \\
		$E_{leak}$ &  \SI{-70.0}{\milli\volt} & \SI{-65.0}{\milli\volt} \\
		$E_{exc}$ & - & \SI{0.0}{\milli\volt} \\
		$E_{inh}$ & - & \SI{-80.0}{\milli\volt} \\
		$E_{th}$ & \SI{-54.0}{\milli\volt} & \SI{-50.0}{\milli\volt} or Eqs. \ref{eq:ad_thr_diff} and \ref{eq:ad_thr} \\
		$\tau_{ref}$ & \SI{1.0}{\milli\second} & \SI{2.0}{\milli\second} \\
		$\tau_{exc}$ & - & \SI{0.5}{\milli\second} \\
		$\tau_{inh}$ & - & \SI{10.0}{\milli\second} \\
		$\tau_{th}$ & - & \num{1.0d4}\si{\second} \\
		$C_{th}$ & - & \SI{20}{\milli\volt\second} \\
		\noalign{\smallskip}\hline
	\end{tabular}
   \label{tab:neuron_parameters}
\end{table}

The excitatory synapses (connecting every sensorial afferent and interneuron pair) were equipped with additive exponential spike-timing dependent plasticity (eSTDP) (Fig. \ref{fig:network-model}A).This unsupervised learning mechanism strengthens the synapses when a presynaptic spike occurs just before a postsynaptic spike. Inversely, when a presynaptic spike happens just after a postsynaptic spike the synapse is weakened. The inhibitory synapses (those connecting the interneurons with each other) implemented symmetric STDP (iSTDP) (Fig. \ref{fig:network-model}A) as previously used \cite{Garrido2016} (see Supplementary Information for a more detailed description of the STDP learning rules). This mechanism increases the weights of those synapses connecting two neurons often firing closely (the firing order is not considered in this case). The temporal parameters of these learning rules were set with the same values as used in \cite{Garrido2016}, while the maximum weight and long-term depression (LTD)/long-term potentiation (LTP) ratio were explored to achieve the best learning performance. 

\subsection{Pattern generation and stimulation protocol}

The sensorial afferent neurons received simultaneously stimulation patterns and background noise simultaneously as current (Fig. \ref{fig:network-model}A). The resulting current was integrated by the afferent neurons and transformed into spike bursts containing between 1 and 3 spikes per oscillatory cycle. It is worth mentioning that in this article a stimulation pattern refers to a combination of input current values conveyed to a set of input (excitatory) neurons. Every time that each pattern was active the same combination of currents stimulated the same selected afferent neurons. The input currents and the afferent neurons involved in each pattern were randomly chosen according to a uniform distribution. Each pattern included only 10 percent of the input neurons (240 over 2400 inputs) and presented 25 percent of the simulation time. 

Aiming to provide a challenging learning task, two important design decisions were taken during the creation of the stimulation currents:
\begin{enumerate}
	\item Up to 4 different (spatially overlapping) patterns for each simulation were presented.
	\item The current activity (including background noise and stimulation patterns) were normalized ensuring that all the neurons received the same average current throughout the simulation and the average current sent to the whole population was the same at any time. It ensured that no clues existed (i.e. different levels of global activity) when a pattern was presented.
\end{enumerate}

The stimulation currents have been calculated for each time bin (average duration 250ms). In addition to these stimuli, an 8-Hz-frequency sinusoidal current has been injected to the input neurons (Fig. \ref{fig:network-model}). The amplitude of this current was set to \num{0.15}\si{\pico\ampere}. This stimulation corresponds to theta-frequency oscillatory activity as has been reported to occur in the cerebellar input layer \cite{Gandolfi2013} and other brain areas \cite{masquelier2009current}. 

The network model simulation was allowed to evolve for \SI{3000}{\second}, time enough for the plasticity mechanisms to converge to a steady state. 

\subsection{Selectivity analysis} \label{subsec:selectivity-analysis}

In order to assess the learning capabilities of the network (including synaptic plasticity) the amount of information transmitted from the input layer to the interneuron population was evaluated. Contrary to our previous research where the interneuron population contained a reduced number of neurons \cite{Garrido2016}, in this case the realistic density of neurons prevented the mutual information of the whole population to be estimated. Thus, when only one input pattern was presented the mutual information ($MI$) between the input layer and each individual interneuron was analyzed. This approach assumes that each interneuron was able to encode the input pattern independently (instead of considering the combinations of active/inactive neurons as was done previously). This assumption was especially restrictive with small fan-in connectivity ratios, where the inputs to a single neuron did not cover all the input fibers involved in a pattern. However, it was a computationally tractable method to assess the learning goodness of a network configuration. The MI between the single neuron activity and the stimulation pattern was calculated according to Eq. \ref{eq:mi}

\begin{equation}
\label{eq:mi}
MI=H(S)+H(R)-H(S,R)
\end{equation}

where $H(S)$ represents the entropy of the stimulus, $H(R)$ the entropy of the neuron activity and $H(S,R)$ the joint entropy of the stimulus and the response. They were defined as follows:

\begin{equation}
\label{eq:hs}
H(S)=-\sum_{s\in S}^{}P(s)\log_{2}(P(s)),
\end{equation}

\begin{equation}
\label{eq:hr}
H(R)=-\sum_{r\in R}^{}P(r)\log_{2}(P(r)),
\end{equation}

\begin{equation}
\label{eq:hsr}
H(S,R)=-\sum_{s\in S}^{}\sum_{r\in R}^{}P(s,r)\log_{2}(P(s,r))
\end{equation}

where $S$ is the set including the two possible states of the input pattern (presence/absence) and $R$ is the two possible states of firing/silence of each inhibitory interneuron. $P$ represents the estimation of the probability of occurrence of the stimulus, the response or both simultaneously. This estimation was calculated considering the final \SI{300}{\second} of the simulation. Finally, the $MI$ was averaged over the whole population of interneurons.

An upper bound of the $MI$ was calculated based on the idea that a perfect detector would present the same entropy in the input as the output and the joint entropy ($H(S)=H(R)=H(S,R)$). In that case, the optimal mutual information would be $MI_{max}=H(S)$. The ratio between the MI and the maximal MI provides a normalized estimator for quantifying the learning capability of the model, according to the uncertainty coefficient ($UC$), defined as follows:

\begin{equation}
\label{eq:uc}
UC=\frac{<MI>}{MI_{max}}=\frac{<H(S)+H(R)-H(S,R)>}{H(S)}
\end{equation}

where $<MI>$ is the $MI$ averaged over the population of interneurons. Therefore, a perfect detector would obtain $UC=1$.

When multiple stimulation patterns were used a simplified approach was followed to assess the learning capability. In this case, the hit-rate of each neuron was calculated in those bins where the pattern was present in the input. Each neuron was assigned as a representative of the pattern where that neuron obtained the highest hit-rate (i.e. the stimulation pattern leading the neuron to the highest activity). Then, the false-alarm rate was calculated with respect to that pattern, and the difference between these two values was used as the estimator of the learning accuracy (LA) of that neuron.

\begin{equation}
\label{eq:la}
LA(n)=\text{hit}(n,p_n)-\text{false-alarm}(n,p_n)
\end{equation}

where $LA(n)$ is the learning accuracy of the interneuron $n$, $\text{hit}(n,p_n)$ is the hit-rate of the neuron $n$ regarding the pattern $p_n$ (the pattern in response to which the neuron is more active) and $\text{false-alarm}(n,p_n)$ represents the false-alarm-rate from the neuron in response to the same pattern. Finally, the average $LA(n)$ index of the whole neuron population was used as an estimator of the learning goodness. An upper bound of this estimator is 1.0, meaning that the neuron fires when, and only when, this particular pattern is present in the input ($\text{hit}=1$ and $\text{false-alarm}=0$).

\subsection{Network parameter optimization with evolutionary algorithms}

Evolutionary algorithms (EAs) represent a powerful tool for optimizing the configuration (parameters) of complex systems. In particular, previous studies have applied EAs to spiking neuronal networks in different ways, adjusting neuron model parameters to reproduce experimentally recorded membrane potential traces \cite{VanGeit2016}, firing properties \cite{Venkadesh2017} or tuning network parameters to reproduce a target population activity \cite{Russell2010}. In order to maximize the transmission of information between the input stimulation and the output activity of the network under study EAs were implemented to explore the parameter space of the STDP learning rules (Fig. \ref{fig:network-model}B). Thus, for every network structure (configuration) under study an EA was run aiming to obtain the best configuration of the learning parameters (i.e. the LTD/LTP ratio and the maximum synaptic weight of the eSTDP and the iSTDP rules). The EA was implemented in Python by using the Distributed Evolutionary Algorithms in Python (DEAP) library \cite{Fortin2012}.

The EA was set with a population size of 42 individuals, randomly chosen for the first generation from a uniform distribution. The EA structure implemented in this work is described in  Figure \ref{fig:network-model}B. Each individual represents a network model with different learning rule parameter values (namely, $MaxWeight_{eSTDP}$, $r_{eSTDP}^{LTD/LTP}$, $MaxWeight_{eSTDP}$ and $r_{eSTDP}^{LTD/LTP}$). The evaluation of the fitness function was conveniently run in parallel between all the available computation nodes, as simulation represents most of the computational load. After each network configuration was allowed to evolve for 3000 seconds, the $UC$ or the $LA$ were calculated (see below). In order to produce a new population, crossover, mutation and selection processes were executed. Each generation used the one-point crossover operator with a probability of 0.70. Each individual was selected for mutation with a probability of 0.15 and, if selected, each parameter was mutated with a probability of 0.1. The selection operator was set to 3-size tournament. The algorithm was left to evolve until the 5 individuals of the hall of fame stayed unchanged for 3 successive generations.

In those EAs aiming to optimize the configuration for detecting only one stimulation pattern the $UC$ was used as the fitness function. However, in the case of multiple stimulation patterns, the fitness function was redesigned in order to obtain different output neuron subpopulations responsive to each stimulation pattern. The subpopulations of those neurons (72/4=18 neurons) with the highest $MI$ for each pattern were selected. The average $MI$ of each neuron with respect to its preferred pattern was used as the fitness function. This function ensures a homogeneous distribution of the output neurons between the corresponding patterns.

\section{Pattern recognition analysis with SimPart and StackViz} \label{sec:4-results}
This section details through a series of experiments how we used the work-flow presented here, and more concretely the visualization tools SimPart and StackViz, for the analysis and validation of the learning performance. In order to evaluate the effectiveness of the work-flow for these purposes, we contrast the numerical results with the visual ones generated by the visualization tools. Firstly we apply our framework to evaluate how neurons respond to a single pattern with both non-adaptive and adaptive firing threshold. The second one analyzes how neurons respond to multiple input patterns also with different firing thresholds. Finally, the third one analyzes the impact of the number of input fibers stimulating the circuit on the effectiveness of the learning mechanisms.

\subsection{Receptive field emergence with a single stimulating pattern and eSTDP}

Aiming to demonstrate the benefits of the proposed visualization framework the experimental results shown in a previous article \cite{masquelier2009current} were first reproduced. This study evidenced how receptive fields emerge in the afferent connections (governed by excitatory STDP) of interneuron populations. Thus, a network configuration including 2000 input fibers per interneuron, a single stimulating pattern and non-adaptive firing threshold in the interneurons was simulated.

\begin{figure*}
	\includegraphics[width=1.00\textwidth]{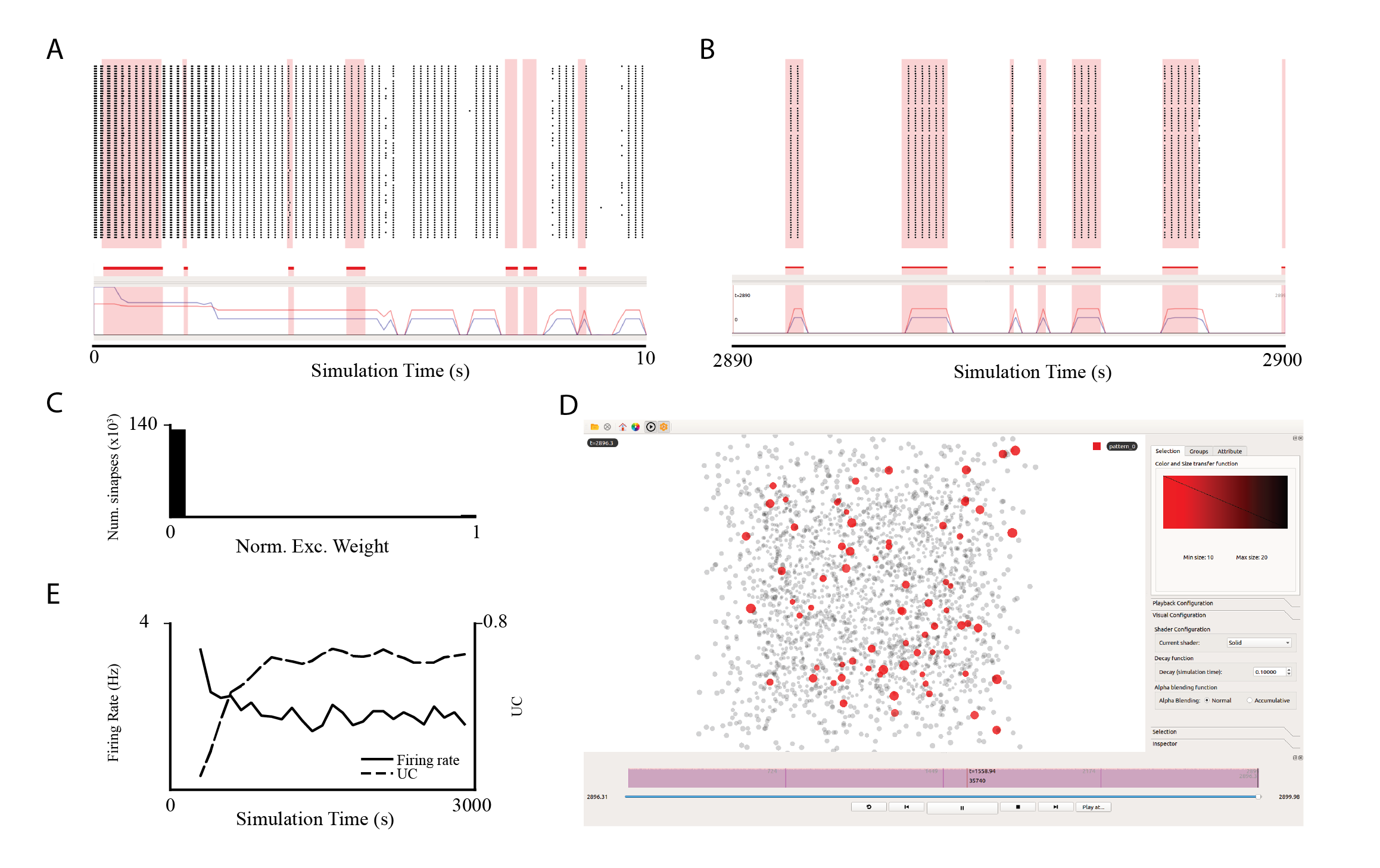}
	\caption{The learning evolution of the network when being presented with a single stimulating pattern. The interneurons have been set with static firing thresholds. (A) (Top) Raster plot of the interneuron population during the initial \SI{10}{\second}. The presence of the stimulating pattern is highlighted with red background. (Bottom) Activity histogram with local (blue solid line in linear scale) and global (red solid line in logarithmic scale) normalization generated with StackViz during the same time. The pattern presentation is also shown as red horizontal bars and red shadows as indicated in StackViz. (B) (Top) Raster plot of the interneuron population during the final \SI{10}{\second} and (Bottom) activity histogram generated with StackViz. (C) Histogram of the excitatory synaptic weights after the learning process. (D) Spatial representation of the network activity generated in SimPart during the presentation of the stimulation pattern at the end of the learning. Note that in this view only the interneurons were selected to be shown. (E) Evolution of the interneuron firing rate (left axis, solid line) and the $UC$ (right axis, dashed line) considering 300-second sliding windows.}
	\label{fig:single_pattern_FT}
\end{figure*}

At the beginning of the simulation, with homogeneous initial synaptic weights (randomly generated according to a uniform distribution in a short range), every interneuron emitted between two and three spikes per oscillatory cycle, regardless of the presence/absence of the stimulation pattern (Fig. \ref{fig:single_pattern_FT}A). This independence of the pattern presence is better observed in the plain activity histograms in StackViz (Fig. \ref{fig:single_pattern_FT}A). However, as the simulation advanced the interneurons became sensitive to the presentation of the pattern, reaching almost perfect recognition at the end of the simulation time (Fig. \ref{fig:single_pattern_FT}B). In this case, the population activity histograms in StackViz produced remarkable peaks when the input pattern was present (Movie S1 in Supplementary Materials).

During the learning process, the eSTDP learning rule drove the synaptic weights to bi-modal distribution with most of the synapses becoming almost zero while a few of them (connecting the most relevant fibers included in the pattern) saturated up to the maximum allowed value (Fig. \ref{fig:single_pattern_FT}C). The spatial representation of the network activity generated in SimPart evidenced the homogeneous distribution of the recruited interneurons (Fig. \ref{fig:single_pattern_FT}D) as was expected from the uniform distribution used to generate both the afferent neurons and the connectivity to the interneurons. The eSTDP learning rule reduced most of the afferent connections, decreasing the firing rate of the output neurons as a consequence (Fig. \ref{fig:single_pattern_FT}E). Simultaneously, the output neurons became responsive to the presence of the pattern, increasing the mutual information as depicted from the evolution of the $UC$ index (Fig. \ref{fig:single_pattern_FT}E).

The highest $UC$ index was found with $LTD/LTP_{ratio}=1.46$ and $MaxWeight_{eSTDP}=$\SI{0.7}{\nano\siemens}. During the evolution of the learning the $UC$ index reached the steady state after \SI{1000}{\second} with values over 0.6 ($UC=0.65$ at the end of the simulation time) (Fig. \ref{fig:single_pattern_FT}E). This represented a very accurate recognition capability.

\begin{figure*}
	\includegraphics[width=1.00\textwidth]{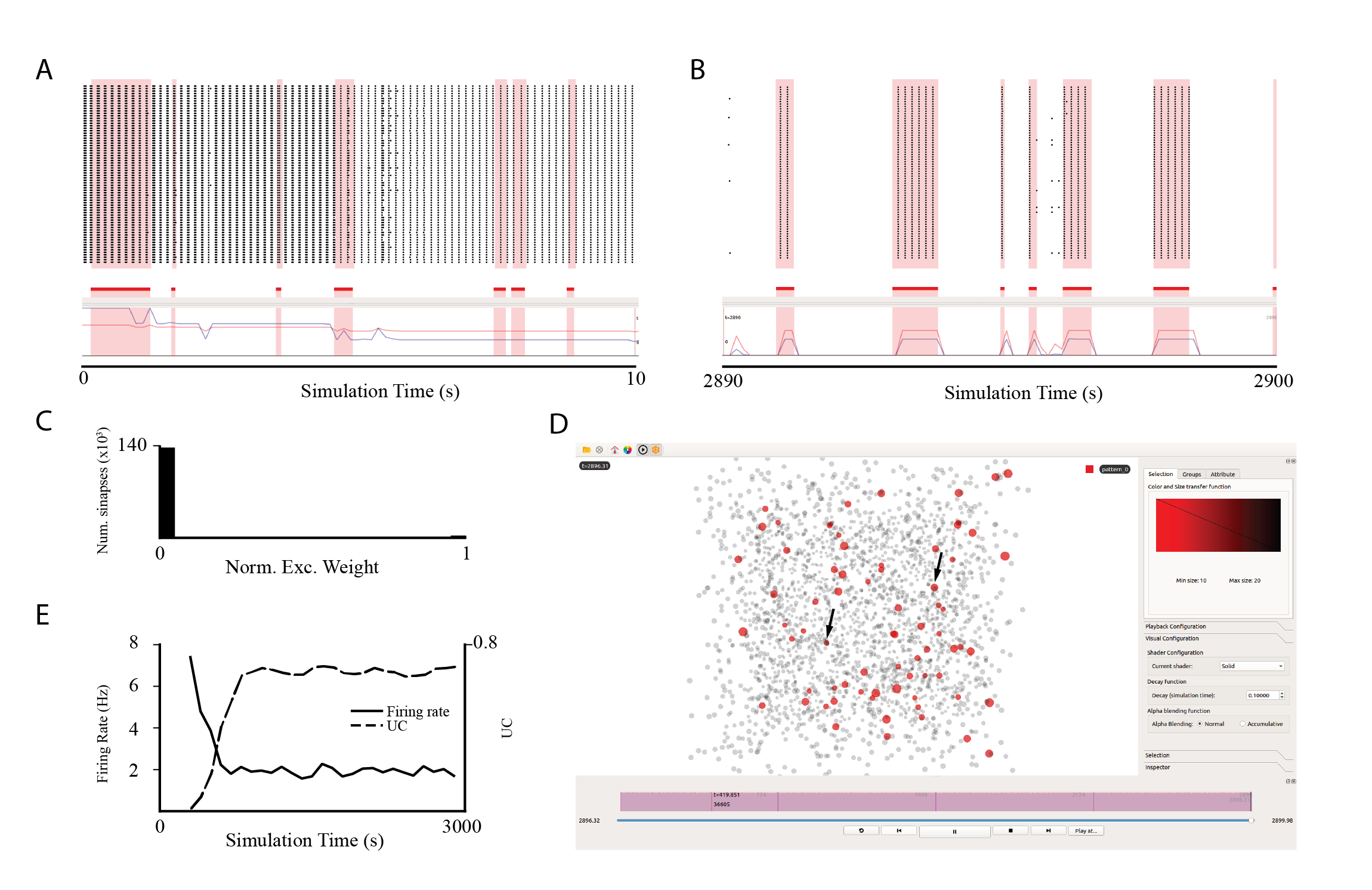}
	\caption{The learning evolution of the network when being excited with a single stimulating pattern. The interneurons have been set with adaptive firing threshold. (A) (Top) Raster plot of the interneuron population during the initial \SI{10}{\second}. The presence of the stimulating pattern is highlighted with red background. (Bottom) Activity histogram with local (blue solid line in linear scale) and global (red solid line in logarithmic scale) normalization generated with StackViz during the same time. (B) (Top) Raster plot of the interneuron population during the final \SI{10}{\second} and (Bottom) activity histogram generated with StackViz. (C) Histogram of the excitatory synaptic weights after the learning process. (D) Spatial representation of the network activity generated in SimPart during the presentation of the stimulation pattern at the end of the learning. Note that in this view only the interneurons were selected to be shown. (E) Evolution of the interneuron firing rate (left axis, solid line) and the $UC$ (right axis, dashed line) considering 300-second sliding windows.}
	\label{fig:single_pattern_AT}
\end{figure*}

Nevertheless, in this case, 2 interneurons became almost silent (missing the pattern) during the simulation (Fig. \ref{fig:single_pattern_FT}B). This issue was corrected by using the adaptive firing threshold mechanism presented in Eq. \ref{eq:ad_thr_diff} and Eq. \ref{eq:ad_thr}. When the interneurons were equipped with an adaptive threshold, and even though they had started with high firing rates (Fig. \ref{fig:single_pattern_AT}A), all the neurons reached similar firing rates (near the target value of \SI{2}{\hertz}) at the end of the simulation (Fig. \ref{fig:single_pattern_AT}B). In this case, there were no silent interneurons. This homeostatic mechanism did not alter the learning capabilities of the system as is shown in the activity histograms generated in StackViz (Fig. \ref{fig:single_pattern_AT}B), the bi-modal weight distribution (Fig. \ref{fig:single_pattern_AT}C), and the spatial representation of the network activity (Fig. \ref{fig:single_pattern_AT}D) obtained with SimPart (Movie S2 in Supplementary Materials).

The SimPart visualization of the activity in the final learning stage evidences the emergence of two additional active interneurons (see black arrows in Figure 5D) when using adaptive threshold in comparison with using fixed threshold. The best network configuration using an adaptive firing threshold was found with $LTD/LTP_{ratio}=1.0$ and $MaxWeight_{eSTDP}=$\SI{1.14}{\nano\siemens}. It obtained a final $UC$ value of 0.69, and the steady state was reached after \SI{1000}{\second} (Fig. \ref{fig:single_pattern_AT}E). Thus, no relevant improvement was observed in the learning capabilities of the network when using adaptive firing threshold if a single stimulation pattern is presented.
    
\subsection{Multiple pattern recognition with eSTDP and iSTDP} \label{subsec:results-multiple-pattern}

Once the network had learned and detected a single pattern, we explored to what extent the system would be able to recognize several input patterns simultaneously. In this framework, it was expected that each interneuron would become sensitive to one pattern. The recognition of multiple patterns required the inclusion of lateral inhibition between the interneurons to avoid all the neurons becoming sensitive to the same pattern (the one dominating the beginning of the simulation). The authors' previous research had shown enhanced recognition capability when using $iSTDP$ instead of fixed lateral inhibition, allowing inhibition to decrease once different sub-populations had become sensitive to different patterns. Thus, after fixing the $eSTDP$ parameter to the best found values for each configuration the parameter space of the $iSTDP$ learning rule was explored.

\begin{figure*}
	\includegraphics[width=1.00\textwidth]{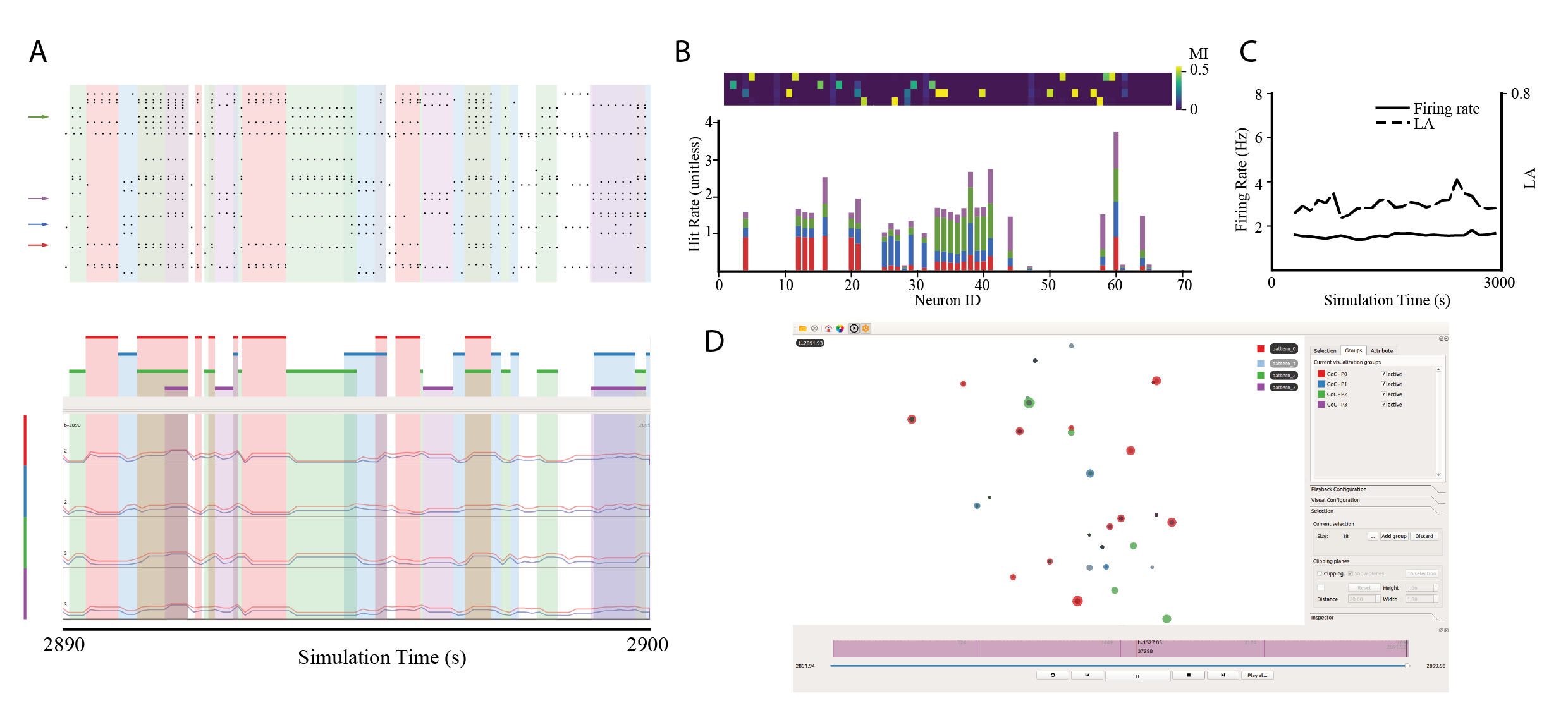}
	\caption{The learning evolution of the network when being presented with 4 partially overlapping stimulating patterns. The interneurons have been equipped with lateral inhibition and static firing threshold (A) (Top) Raster plot of the interneuron population during the final \SI{10}{\second}. The color arrows indicate the activity of four neurons responding to each one of the patterns. (Bottom) Activity histogram obtained from the StackViz application. The presence of the stimulating patterns is highlighted with red, blue, green and violet backgrounds (and color combinations indicating the presentation of more than one). Each histogram represents the activity of a different neuron sub-population with the highest $MI$ with each stimulation pattern (as indicated in the vertical bars to the left) (B) (Top) $MI$ matrix representing the MI between each neuron (column) and each stimulation pattern (row). (Bottom) Stacked bar plot of the hit-rate between each neuron and each stimulation pattern. Note that in this case the neurons have been sorted according to the preferred pattern. (C) Evolution of the interneuron firing rate (left axis, solid line) and the average $LA(n)$ (right axis, dashed line) considering 300-second sliding-windows. (D) SimPart view of the network at time $t=2891.8s$. Each neuron has been highlighted with the color of its preferred pattern. Note that, at this time stamp, three different patterns (red, green and violet) are active as shown in the color legend.}
	\label{fig:multiple_pattern_FT}
\end{figure*}

The simulations showed that, at the end of the simulation period, some of the neurons become sensitive (and fired in response) to different patterns (Fig. \ref{fig:multiple_pattern_FT}A). This example evidenced the automatic grouping functionality implemented in SimPart. Considering the $MI$ between each of the stimulation patterns and the activity of each neuron, the software automatically grouped those neurons which were more responsive to each of the 4 presented patterns and generated the corresponding activity histograms (Fig. \ref{fig:multiple_pattern_FT}A). The hit-rate stacked histogram and the MI matrix showed that a variable number of neurons fired in response to the presence of their preferred patterns. This number ranged from 4 neurons in response to the violet pattern to up to 10 neurons in response to the green pattern (Fig. \ref{fig:multiple_pattern_FT}B and Movie S3 in Supplementary Materials).

However, when using the constant firing threshold, many of the neurons became almost silent after the learning process (Fig. \ref{fig:multiple_pattern_FT}A), obtaining extremely low $MI$ (Fig. \ref{fig:multiple_pattern_FT}B). The silent neurons resulted in low average $LA(n)$, indicating that the difference between firing to preferred and non-preferred patterns was reduced as shown in Figure \ref{fig:multiple_pattern_FT}C. The spatial view generated by SimPart also evidenced the heterogeneous response of the different neurons to the presentation of the patterns (Fig. \ref{fig:multiple_pattern_FT}D).

\begin{figure*}
	\includegraphics[width=1.00\textwidth]{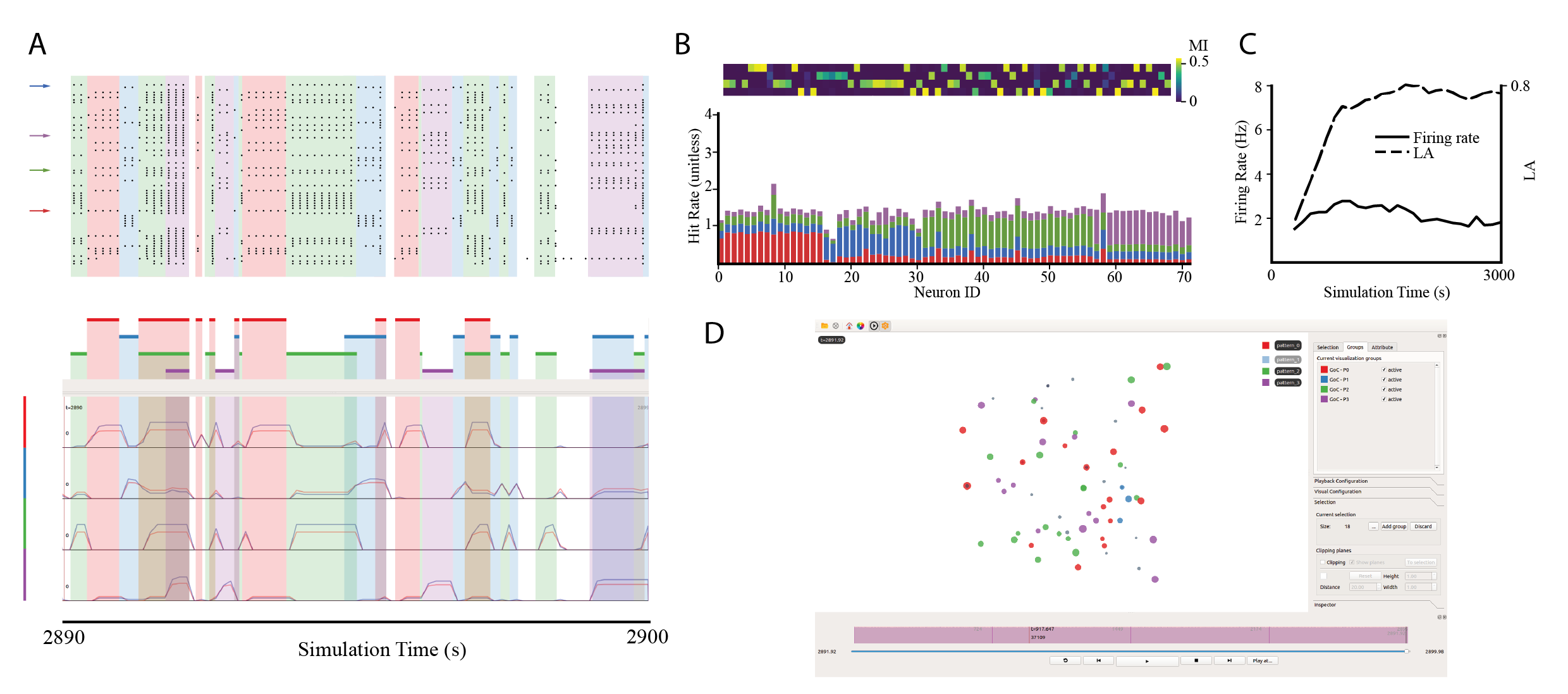}
	\caption{Pattern recognition with 4 partially overlapping stimulating patterns and adaptive firing threshold (A) (Top) Raster plot of the interneuron population during \SI{10}{\second} at the end of the learning process. The color arrows indicate the activity of four neurons responding to each one of the patterns. (Bottom) Activity histogram generated by the StackViz application. The presence of the stimulating patterns is highlighted with red, blue, green and violet background (and color combinations indicating the presentation of more than one). Each histogram represents the activity of a different neuron sub-population with the highest $MI$ with each stimulation pattern (as indicated in the vertical bars to the left) (B) (Top) $MI$ matrix representing the MI between each neuron (column) and each stimulation pattern (row). (Bottom) Stacked bar plot of the hit-rate between each neuron and each stimulation pattern. Note that in this case the neurons have been sorted according to the preferred pattern. (C) Evolution of the interneuron firing rate (left axis, solid line) and the average $LA(n)$ (right axis, dashed line) considering 300-second sliding-windows. (D) SimPart view of the network at time $t=2891.79s$. Each neuron has been highlighted with the color of its preferred pattern. Note that, at this time stamp, three different patterns (red, green and violet) are active as shown in the color legend.}
	\label{fig:multiple_pattern_AT}
\end{figure*}

The previous limitations were partially overcome when using the adaptive threshold mechanism, which facilitated the development of homogeneous sub-populations of neurons responsive to each of the stimulating pattern (Fig. \ref{fig:multiple_pattern_AT}A) as can be observed in the SimPart histograms (Fig. \ref{fig:multiple_pattern_AT}A). By reducing the number of silent neurons, the adaptive threshold mechanism increased the number of interneurons dedicated to the detection of each pattern (Fig. \ref{fig:multiple_pattern_AT}B). Thus, the number of neurons that fired in response to each pattern ranged between 15 (blue and violet patterns) and 26 (green pattern) neurons. This enhanced pattern representation resulted in a higher average $LA(n)$ index and firing rate (Fig. \ref{fig:multiple_pattern_AT}C). The SimPart network representation allowed easy identification of active neurons in the groups corresponding to each active pattern (Fig. \ref{fig:multiple_pattern_AT}D and Movie S4 in Supplementary Materials).

\subsection{Dependence on the number of input fibers}

\begin{figure*}
	\includegraphics[width=1.00\textwidth]{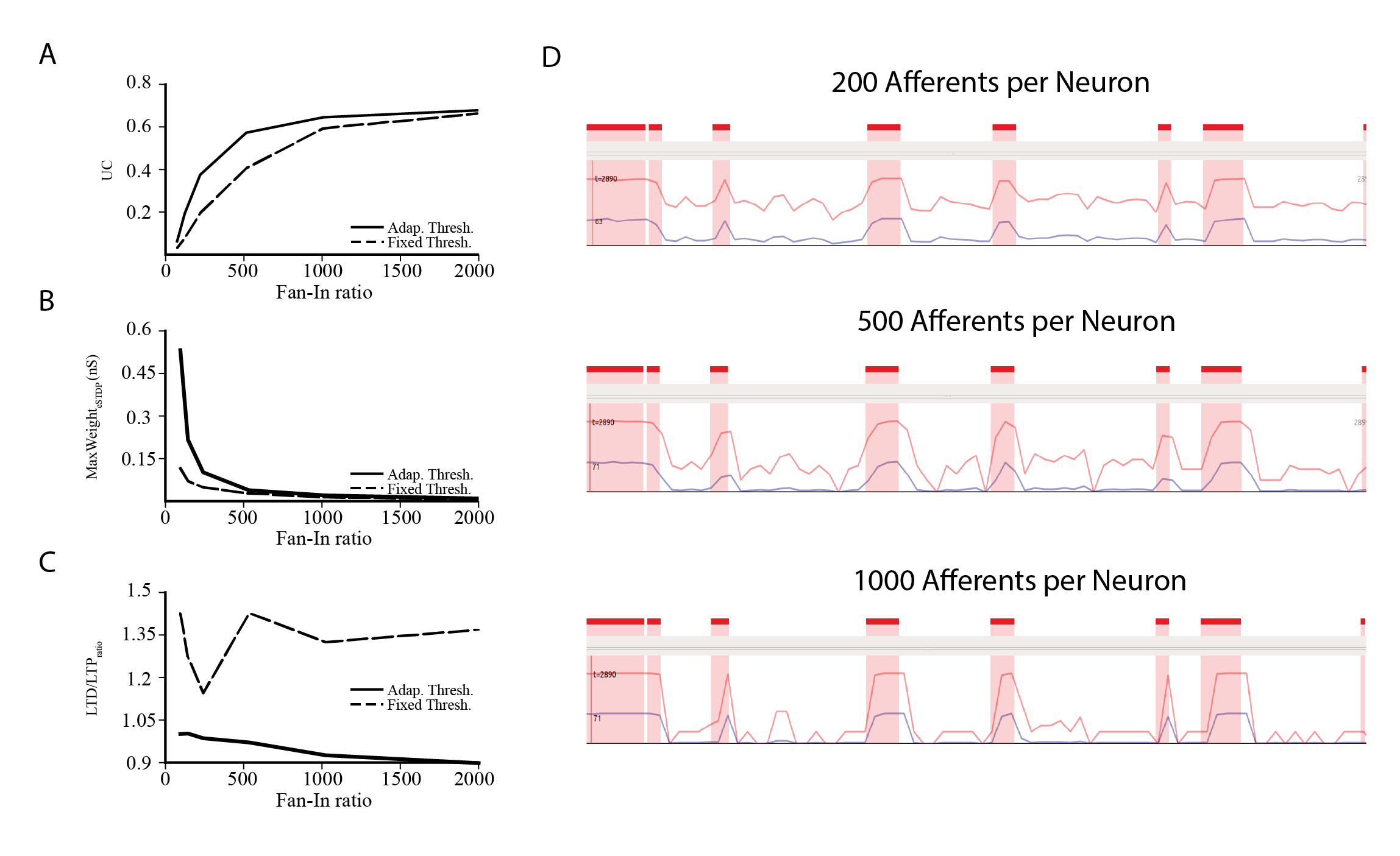}
	\caption{Learning accuracy exploration with single stimulating pattern and different fan-in ratios. (A) $UC$ obtained by the EA in the best individual with different fan-in ratios. (B) $MaxWeight_{eSTDP}$ and $LTD/LTP_{ratio}$ values of the best configuration. (C) StackViz representation of the activity histogram at the end of learning with 200, 500 and 1000 sensorial afferents per inhibitory interneuron. Note that, for these simulations, the interneurons were equipped with an adaptive threshold.}
	\label{fig:fan_in_ratio}
\end{figure*}

Since the information transfer between each stimulation pattern and every individual interneuron was analyzed, it was expected that the pattern detection capability might decrease as a consequence of the reduction in the number of input fibers reaching each interneuron. Thus, how the system degraded as the number of input fibers fell was quantified. The best parameter configuration was obtained for different network settings accounting with fan-in ratio between 50 and 2000 sensorial afferents per inhibitory interneuron. Note that 2,400 inputs per neuron would be needed to allow them a full view of the stimulation patterns. As expected, the $UC$ showed remarkabe degradation as the fan-in ratio decreases (from near $UC=0.7$ with 2000 inputs per interneuron to $UC=0.05$ with only 50 inputs per interneuron). It is relevant to note that the exploration evidenced that the learning capability degradation occurred mainly below 500 inputs per neuron (Fig. \ref{fig:fan_in_ratio}A) and followed logarithmic evolution with the number of inputs.

The use of an adaptive threshold turned out to be remarkably beneficial with fan-in ratios around the cut-off values (between 200 and 500 afferents per interneuron) (Fig. \ref{fig:fan_in_ratio}A and Movies S5, S6 and S7 in Supplementary Materials). The $UC$ obtained with adaptive threshold ($UC_{200}$=0.38 and $UC_{500}$=0.57 with 200 and 500 afferents respectively) notably outperformed the $UC$ values obtained in absence of this mechanism ($UC_{200}$=0.20 and $UC_{500}$=0.41). In these cases, the adaptive threshold mechanisms effectively controlled the firing rate, avoiding the emergence of silent neurons during the learning process and allowed these neurons to become responsive to the presented pattern, thus increasing the $MI$.

The $MaxWeight_{eSTDP}$ parameters optimizing each network configuration followed exponential decay (Fig. \ref{fig:fan_in_ratio}D) with the fan-in ratio. Not unexpectedly, as the number of afferents per interneuron was reduced to half the maximum excitatory weight of the learning rule had to be duplicated to compensate the reduction in the number of inputs per interneuron. The $LTD/LTP_{ratio}$ parameter also followed linear decay in the case of adaptive threshold. A possible explanation for this fact is that the $LTD/LTP_{ratio}$ was associated with the learning speed, indicating that as the number of afferents was decreased the learning speed had to be set more aggressively. In the network with a fixed threshold, however, the $LTD/LTP_{ratio}$ did not seem to vary as a function of the fan-in ratio (Fig. \ref{fig:fan_in_ratio}C and Movie S5 -200 input synapses per neuron-, S6 -500 input synapses per neuron- and S7 -1000 input synapses per neuron- in Supplementary Materials).

The enhanced learning capabilities as a function of the fan-in ratio were especially visible when comparing the activity histograms generated by StackViz (Fig. \ref{fig:fan_in_ratio}D). While only 63 of the 72 interneurons became active on response to prolonged pattern presentation in the 200-afferent-per-neuron configuration (top), 71 of the 72 interneurons became active when connecting 500 (medium) or 1000 (bottom) afferents to each interneuron. Additionally, the absence of activity in the pattern absence periods benefited from the increased number of afferents (Fig. \ref{fig:fan_in_ratio}). Based on these results, reliable pattern recognition in a converging network with eSTDP and adaptive threshold requires, at least, 500 input afferents per target neuron. This number increased to 1000 afferents when using a fixed firing threshold.

\section{Conclusion and discussion}\label{sec:5-discussion}
The simulations presented in this paper represent a use case that validates the integration of generic visualization tools (such as  SimPart and StackViz), together with additional problem-specific optimization and analysis custom scripts in order to study the evolution of learning systems and the influence of particular network configurations. 

This theoretical study of the inhibitory interneuron learning evidences how interactive simulation activity playback represents a powerful tool for studying systems involving oscillations (such as the network proposed here) and the synchronization/precision of spike emission.

The disaggregation capabilities of both StackViz and SimPark provide a fast and accurate way of analyzing arbitrary subset activity and assessing the responsiveness of neurons to the presented input patterns. This is performed in the former by the enhanced plot stacking features that display the pattern events in a simple but powerful representation; and in the latter through the use of visual differentiation of functional neuron groups. These features allow the user to highlight neuronal activity with the highest response rate to specific patterns. In learning systems, such as the one proposed here, this representation provides a preliminary evaluation of the information transmission between the input and the output layers.

Interactive visualization facilitated the interpretation of the different results and specific set ups at several stages of the study (e.g, preliminary simulation set-up, result extraction, validation of the best configurations and learning convergence). Using all these tools together remarkably facilitated the analysis and configuration of a complex learning system such as the use-case presented here. Moreover, the presented work-flow does opens up a wide range of possibilities for testing spiking-neuronal-network-based learning systems, beyond the parameter optimization of the learning system. Our work-flow includes an additional stage of enhanced visual validation. Thus, it could be useful for developing improved mechanisms of the learning system (e.g., neuron and synaptic models) or even improving the EA itself (e.g., refining the fitness function with more accurate learning performance estimators, adjusting EA parameters and testing novel EAs).

The simulations show that, in the proposed framework, the fan-in ratio required to correctly learn and recognize input patterns is around 500 inputs per interneuron. This number is in agreement with previous simulation results where the proportion of pattern-carrying afferents required for correct learning falls over 480 mossy fibers per pattern\cite{masquelier2009current}. However, since the number of afferents per neuron has been studied rather than the ratio of inputs included in the pattern, these results demonstrate that restricted input connectivity may also be effective for allowing the eSTDP mechanism to extract the most relevant inputs, thus leading to reduced connectivity and faster simulations with similar learning capabilities.

The presented learning structured could be analogous to several biological areas, such as the cerebellar cortex, the insect mushroom body, and dentate gyrus \cite{Cayco2019}. The simulation system we describe follows the histological data of the cerebellum, taken into account both the density of fibers/neurons of two of the main elements in the input layer: the mossy fibers (sensorial afferents) and the Golgi cells (inhibitory interneurons). In addition to this, there is evidence of long-term plasticity at different synaptic sites in the cerebellar granular layer \cite{Sgritta2017,Mapelli2016}. Our results suggest that the fan-in connectivity ratio existing in the mossy fiber-Golgi cell connections, rather than reduced mossy fiber-granule cell connections, could support sensorial pattern recognition. The cerebellar input layer interneurons (the Golgi cells) receive, on average, 228 sensorial afferent fibers \cite{Pellionisz1973}. Our simulations indicate that additional mechanisms (beyond feedforward eSTDP) are required in order to achieve accurate pattern recognition. The feedback input from the granule cells (the excitatory layer being the output of the granular layer) could carry out this task as each interneuron receives input from almost 5000 granule cells \cite{Pellionisz1973}.

The proposed learning environment also gives a role to theta-frequency band oscillations in allowing the output neurons to learn current-based input patterns. Intrinsic neuron properties, such as the firing resonance in the same frequency-band have been experimentally validated \cite{Solinas2007} and may enhance the learning properties of the proposed system. In any case, further exploration considering the particularities of the cerebellar system seems to be required and the proposed visualization framework represents a helpful tool to explore the interplay between STDP learning, oscillations and locality constraints of the inputs within a neuronal system simulation case study.

\section*{Acknowledgments}
The simulations have been run in the ALHAMBRA supercomputer of
the University of Granada.
Jes\'us Garrido is supported by the University of Granada under the
Young Researchers Fellowship, by the Operative Program FEDER 2014-2020 and by the Ministry of Economy, Knowledge and Innovation of the Andalusian Regional Government under the EmbBrain (A-TIC-276-UGR18) project. Additionally, this research has received funding from the Spanish Ministry of Economy and Competitiveness under grants MINECO-FEDER TIN2016-81041-R, C080020-09 (Cajal Blue Brain Project, Spanish partner of the Blue Brain Project initiative from EPFL), TIN2014-57481 and TIN2017-83132 as well as from the European Union's Horizon 2020 Framework Program for Research and Innovation under the Specific Grant Agreement No. 785907 (Human Brain Project SGA2).

\bibliographystyle{elsarticle-num}

\bibliography{biblio}

\section*{Supplementary Information}
\subsection*{The neuron and synapse models}

The excitatory afferent neurons were modeled using a current-based version of the LIF. In this neuron model the membrane potential ($V_m$) is computed through the following differential equation:

\begin{equation}
\label{eq:vm_cur}
\frac{dV_{m}}{dt} = -\frac{V_{m}-E_{rest}}{\tau_{m}}+\frac{I_{e}(t)}{C_{m}}
\end{equation}

where $C_m$ denotes the membrane capacitance, $V_m$ is the membrane potential, $E_{rest}$ represents the resting potential, $\tau_m$ is the membrane time constant and finally, $I_e(t)$ is the total excitatory current stimulating the neuron at a certain time $t$. These constants have been set as indicated in Table \ref{tab:neuron_parameters}. Every time the membrane potential overpasses the threshold potential ($V_{th}$) a spike is elicited and the membrane potential is reset to $E_{rest}$.

The inhibitory interneurons were modeled using a conductance-based version of the LIF model. The membrane potential evolution of this neuron model is described via the following equation:

\begin{equation}
\label{eq:vm_cond}
\frac{dV_{m}}{dt} = \frac{(-g_{leak}\cdot(V_{m}-E_{rest})+I_{syn})}{C_{m}}
\end{equation}

where $V_m$ represents the membrane potential, $E_{rest}$ is the resting potential, $C_m$ and $g_{leak}$ represent, respectively, the membrane capacitance and the leak conductance and $I_{syn}$ is the total current that the neuron receives through the synapses according to the following equation:

\begin{equation}
\label{eq:i_syn}
I_{syn} = -g_{exc}\cdot(V_{m}-E_{exc})-g_{inh}\cdot(V_{m}-E_{inh})
\end{equation}

$E_{exc}$ and $E_{inh}$ represent the reversal potential of the excitatory and inhibitory synapses (Table \ref{tab:neuron_parameters}). When the membrane potential goes above the threshold potential (either fixed or adaptive according to Equations \ref{eq:ad_thr_diff} and \ref{eq:ad_thr} in the main manuscript) a spike is elicited and the membrane potential is reset to $E_{rest}$. The excitatory ($g_{exc}$) and inhibitory ($g_{inh}$) conductance of a particular neuron have been modeled using exponential functions as follows:

\begin{equation}
\label{eq:cond_exc}
\tau_{exc}\cdot\frac{dg_{exc,i}}{dt}=-g_{exc,i}+\sum_{k\in ExcSp_{i}}{w_{j,i}(t_{k})\cdot\delta(t-t_{k})}
\end{equation}
\begin{equation}
\label{eq:cond_inh}
\tau_{inh}\cdot\frac{dg_{inh,i}}{dt}=-g_{inh,i}+\sum_{k\in InhSp_{i}}{w_{j,i}(t_{k})\cdot\delta(t-t_{k})}
\end{equation}

where $\tau_{exc}$ and $\tau_{inh}$ represent the excitatory and inhibitory time constants (Table \ref{tab:neuron_parameters}), $ExcSp_{i}$/$InhSp_{i}$ is the set of spikes reaching the neuron $i$ through the excitatory/inhibitory afferent synapses and $\delta(t)$ is the Dirac delta function. According to these equations, every time a spike is received (at time $t_k$) through an excitatory or inhibitory connection (linking the presynaptic neuron $j$ and the postsynaptic neuron $i$), the excitatory or inhibitory conductance is increased accordingly to the synaptic weight existing in that synapse ($w_{j,i}$). For simplicity, these equations consider both excitatory and inhibitory conductance and synaptic weights to be positive or zero.

\subsection*{Spike-Time dependent plasticity}

The weights of the excitatory synapses (those connecting the stimulation fibers with the inhibitory neurons) have been implemented following classic additive excitatory spike-time dependent plasticity (eSTDP). According to this type of Hebbian plasticity, long-term potentiation (LTP) is produced when a postsynaptic spike is elicited shortly after a presynaptic spike. Inversely, a presynaptic spike following a postsynaptic spike will generate long-term depression (LTD) (Fig. \ref{fig:network-model} in the main manuscript). The equations governing the weight change are as follows:

\begin{equation}
\label{eq:estdp}
\Delta w= \begin{cases} A_{eSTDP}^{LTP}\cdot e^{-\frac{t_{post}-t_{pre}}{\tau_{eSTDP}^{+}}} & \mbox{  if } t_{post}\geq t_{pre},\\ A_{eSTDP}^{LTD}\cdot e^{-\frac{t_{pre}-t_{post}}{\tau_{eSTDP}^{-}}} & \mbox{  otherwise }, \end{cases}
\end{equation}

where $\tau_{eSTDP}^{+}$ and $\tau_{eSTDP}^{-}$ are the time constants of the potentiation and depression components and have been set to 16.8ms and 33.7ms respectively. The $t_{post}$ and $t_{pre}$ variables refer to the time when the postsynaptic and presynaptic spikes happen. An all-to-all implementation of this eSTDP has been used, so that all the previous spikes have been considered in order to calculate the total amount of LTD and LTP produced. The maximum amount of LTP ($A_{eSTDP}^{LTP}$) being produced after two (one presynaptic and one postsynaptic) coincident spikes has been set to $A_{eSTDP}^{LTP}=3\cdot 10^{-3}\cdot MaxWeight_{eSTDP}$ while the maximum amount of LTD ($A_{eSTDP}^{LTD}$) has been adjusted in relation to the amount of LTP according to $A_{eSTDP}^{LTD}=r_{eSTDP}^{LTD/LTP}\cdot A_{eSTDP}^{LTP}$. The influence of $MaxWeight_{eSTDP}$ and $r_{eSTDP}^{LTD/LTP}$ parameters have been analyzed throughout this article in different conditions.

\section*{Supplementary Movies}
GoC2000FT.mp4.
Movie S1. Visualization of the learning evolution of the network when being presented with a single stimulating pattern and 2000 input synapses per target neuron. The interneurons have been set with static firing thresholds. The network activity is shown as visualized with the StackViz (left) and ViSimpl (right) tools at the beginning (labelled as before learning) and the ending (labelled as after learning) of the simulation time. See Fig. \ref{fig:single_pattern_FT} for a detailed explanation of the different visual components included in the screen.

GoC2000AT.mp4.
Movie S2. Visualization of the learning evolution of the network when being presented with a single stimulating pattern and 2000 input synapses per target neuron. The interneurons have been set with adaptive firing thresholds. The network activity is shown as visualized with the StackViz (left) and ViSimpl (right) tools at the beginning (labeled as before learning) and the ending (labelled as after learning) of the simulation time. See Figure \ref{fig:single_pattern_AT} for a more detailed explanation of the different visual components included in the screen.

GoC2000FTMulti.mp4.
Movie S3. Visualization of the learning evolution of the network when being presented with four random and overlapping stimulating pattern and 2000 input synapses per target neuron. The interneurons have been set with static firing thresholds. The network activity is shown as visualized with the StackViz (left) and ViSimpl (right) tools at the beginning (labeled as before learning) and the ending (labeled as after learning) of the simulation time. See Figure \ref{fig:multiple_pattern_FT} for a more detailed explanation of the different visual components included in the screen. 

GoC2000ATMulti.mp4.
Movie S4. Visualization of the learning evolution of the network when being presented with a single stimulating pattern and 2000 input synapses per target neuron. The interneurons have been set with adaptive firing thresholds. The network activity is shown as visualized with the StackViz (left) and ViSimpl (right) tools at the beginning (labeled as before learning) and the ending (labeled as after learning) of the simulation time. See Figure \ref{fig:multiple_pattern_AT} for a more detailed explanation of the different visual components included in the screen. 

GoC200FT.mp4.
Movie S5. Visualization of the learning evolution of the network when being presented with a single stimulating pattern and 200 input synapses per target neuron. The interneurons have been set with adaptive firing thresholds. The network activity is shown as visualized with the StackViz (left) and ViSimpl (right) tools at the beginning (labeled as before learning) and the ending (labeled as after learning) of the simulation time. See Figure \ref{fig:fan_in_ratio} for further details of the dataset visualized in this movie.

GoC500FT.mp4
Movie S6. Visualization of the learning evolution of the network when being presented with a single stimulating pattern and 500 input synapses per target neuron. The interneurons have been set with adaptive firing thresholds. The network activity is shown as visualized with the StackViz (left) and ViSimpl (right) tools at the beginning (labeled as before learning) and the ending (labeled as after learning) of the simulation time. See Figure \ref{fig:fan_in_ratio} for further details of the dataset visualized in this movie.

GoC1000FT.mp4
Movie S7. Visualization of the learning evolution of the network when being presented with a single stimulating pattern and 1000 input synapses per target neuron. The interneurons have been set with adaptive firing thresholds. The network activity is shown as visualized with the StackViz (left) and ViSimpl (right) tools at the beginning (labeled as before learning) and the ending (labeled as after learning) of the simulation time. See Figure \ref{fig:fan_in_ratio} for further details of the dataset visualized in this movie.



\end{document}